\PassOptionsToPackage{unicode}{hyperref}
\PassOptionsToPackage{hyphens}{url}
\PassOptionsToPackage{dvipsnames,svgnames,x11names}{xcolor}
\documentclass[
]{article}

\usepackage{amsmath,amssymb}
\usepackage{iftex}
\ifPDFTeX
  \usepackage[T1]{fontenc}
  \usepackage[utf8]{inputenc}
  \usepackage{textcomp} 
\else 
  \usepackage{unicode-math}
  \defaultfontfeatures{Scale=MatchLowercase}
  \defaultfontfeatures[\rmfamily]{Ligatures=TeX,Scale=1}
\fi
\usepackage{lmodern}
\ifPDFTeX\else  
\fi
\IfFileExists{upquote.sty}{\usepackage{upquote}}{}
\IfFileExists{microtype.sty}{
  \usepackage[]{microtype}
  \UseMicrotypeSet[protrusion]{basicmath} 
}{}
\makeatletter
\@ifundefined{KOMAClassName}{
  \IfFileExists{parskip.sty}{%
    \usepackage{parskip}
  }{
    \setlength{\parindent}{0pt}
    \setlength{\parskip}{6pt plus 2pt minus 1pt}}
}{
  \KOMAoptions{parskip=half}}
\makeatother
\usepackage{xcolor}
\usepackage[margin=1in]{geometry}
\makeatletter
\ifx\paragraph\undefined\else
  \let\oldparagraph\paragraph
  \renewcommand{\paragraph}{
    \@ifstar
      \xxxParagraphStar
      \xxxParagraphNoStar
  }
  \newcommand{\xxxParagraphStar}[1]{\oldparagraph*{#1}\mbox{}}
  \newcommand{\xxxParagraphNoStar}[1]{\oldparagraph{#1}\mbox{}}
\fi
\ifx\subparagraph\undefined\else
  \let\oldsubparagraph\subparagraph
  \renewcommand{\subparagraph}{
    \@ifstar
      \xxxSubParagraphStar
      \xxxSubParagraphNoStar
  }
  \newcommand{\xxxSubParagraphStar}[1]{\oldsubparagraph*{#1}\mbox{}}
  \newcommand{\xxxSubParagraphNoStar}[1]{\oldsubparagraph{#1}\mbox{}}
\fi
\makeatother

\usepackage{longtable,booktabs,array}
\usepackage{calc} 
\usepackage{etoolbox}
\makeatletter
\patchcmd\longtable{\par}{\if@noskipsec\mbox{}\fi\par}{}{}
\makeatother
\IfFileExists{footnotehyper.sty}{\usepackage{footnotehyper}}{\usepackage{footnote}}
\makesavenoteenv{longtable}
\usepackage{graphicx}
\makeatletter
\newsavebox\pandoc@box
\newcommand*\pandocbounded[1]{
  \sbox\pandoc@box{#1}%
  \Gscale@div\@tempa{\textheight}{\dimexpr\ht\pandoc@box+\dp\pandoc@box\relax}%
  \Gscale@div\@tempb{\linewidth}{\wd\pandoc@box}%
  \ifdim\@tempb\p@<\@tempa\p@\let\@tempa\@tempb\fi
  \ifdim\@tempa\p@<\p@\scalebox{\@tempa}{\usebox\pandoc@box}%
  \else\usebox{\pandoc@box}%
  \fi%
}
\def\fps@figure{htbp}
\makeatother
\NewDocumentCommand\citeproctext{}{}

\makeatletter
 \let\@cite@ofmt\@firstofone
 \def\@biblabel#1{}
 \def\@cite#1#2{{#1\if@tempswa , #2\fi}}
\makeatother
\newlength{\cslhangindent}
\newlength{\csllabelwidth}
\newenvironment{CSLReferences}[2] 
 {\begin{list}{}{%
  \setlength{\itemindent}{0pt}
  \setlength{\leftmargin}{0pt}
  \setlength{\parsep}{0pt}
  \ifodd #1
   \setlength{\leftmargin}{\cslhangindent}
   \setlength{\itemindent}{-1\cslhangindent}
  \fi
  \setlength{\itemsep}{#2\baselineskip}}}
 {\end{list}}
\usepackage{calc}

\usepackage{booktabs}
\usepackage{longtable}
\usepackage{array}
\usepackage{multirow}
\usepackage{wrapfig}
\usepackage{float}
\usepackage{colortbl}
\usepackage{pdflscape}
\usepackage{tabu}
\usepackage{threeparttable}
\usepackage{threeparttablex}
\usepackage[normalem]{ulem}
\usepackage{makecell}
\usepackage{xcolor}
\makeatletter
\@ifpackageloaded{caption}{}{\usepackage{caption}}
\AtBeginDocument{%
\ifdefined\contentsname
  \renewcommand*\contentsname{Table of contents}
\else
  \newcommand\contentsname{Table of contents}
\fi
\ifdefined\listfigurename
  \renewcommand*\listfigurename{List of Figures}
\else
  \newcommand\listfigurename{List of Figures}
\fi
\ifdefined\listtablename
  \renewcommand*\listtablename{List of Tables}
\else
  \newcommand\listtablename{List of Tables}
\fi
\ifdefined\figurename
  \renewcommand*\figurename{Figure}
\else
  \newcommand\figurename{Figure}
\fi
\ifdefined\tablename
  \renewcommand*\tablename{Table}
\else
  \newcommand\tablename{Table}
\fi
}
\@ifpackageloaded{float}{}{\usepackage{float}}
\floatstyle{ruled}
\@ifundefined{c@chapter}{\newfloat{codelisting}{h}{lop}}{\newfloat{codelisting}{h}{lop}[chapter]}
\floatname{codelisting}{Listing}

\makeatother
\makeatletter
\@ifpackageloaded{caption}{}{\usepackage{caption}}
\@ifpackageloaded{subcaption}{}{\usepackage{subcaption}}
\makeatother

\ifLuaTeX
\usepackage[bidi=basic]{babel}
\else
\usepackage[bidi=default]{babel}
\fi
\babelprovide[main,import]{english}

\def\languageshorthands#1{}
\ifLuaTeX
  \usepackage[english]{selnolig} 
\fi
\usepackage{bookmark}

\IfFileExists{xurl.sty}{\usepackage{xurl}}{} 
\hypersetup{
  pdftitle={Governing Delegation to Generative Artificial Intelligence},
  pdfauthor={Jorge Fábrega},
  pdflang={en},
  colorlinks=true,
  linkcolor={blue},
  filecolor={Maroon},
  citecolor={Blue},
  urlcolor={Blue},
  pdfcreator={LaTeX via pandoc}}

\title{Governing Delegation to Generative Artificial Intelligence}
\usepackage{etoolbox}
\makeatletter
\providecommand{\subtitle}[1]{
  \apptocmd{\@title}{\par {\large #1 \par}}{}{}
}
\makeatother
\subtitle{Human Direction, Work-Related Orientation, and Modes of Use}
\author{Jorge Fábrega (CICS-UDD, Chile)}
\date{2026-08-17}

\begin{document}
\maketitle

\section*{Abstract}\label{abstract}
\addcontentsline{toc}{section}{Abstract}

Delegating cognitive operations to generative artificial intelligence
redistributes execution and raises a governance problem: where human
direction of the task remains. We distinguish two routes. Specified
delegation places that direction before execution, through instructions,
constraints, or criteria that delimit the task. Iterative coproduction
places it during production, through interventions that correct or
redirect provisional outputs. To examine both routes, we use aggregate
monthly cells from the \emph{Anthropic Economic Index} for April and May
2026. The AEI distinguishes two modes of use: 1P API, which corresponds
to direct traffic through Anthropic's API, and Claude.ai, which combines
activity from Chat and Cowork. On this basis, we test whether a stronger
work-related orientation of human-AI interaction is associated with more
specified delegation within each mode and whether the increase in the
iterative profile is greater in Claude.ai than in 1P API. The main
analysis uses level-0 O*NET tasks and estimates how both profiles change
when an eligible record reallocates ten percentage points from personal
use to work-related use. The iterative comparison is restricted to 1,411
node-month pairs observed and eligible in both modes. Specified
delegation increases by 2.76 points in 1P API (95\% CI: {[}2.30,
3.22{]}) and by 1.45 in Claude.ai (95\% CI: {[}0.93, 1.97{]}). On the
common support, iterative coproduction changes by -0.30 points in 1P API
and by 0.15 in Claude.ai, yielding a between-mode difference of 0.45
points (95\% CI: {[}0.15, 0.75{]}). These findings show that
work-related orientation is associated with stronger traces of prior
human direction and that the observable iterative response varies across
modes of use. The article shifts attention from how much the AI executes
to when human direction leaves observable traces.

\textbf{Keywords:} generative artificial intelligence; AI governance;
delegation; 1P API; Claude.ai; compositional data; human-AI
collaboration.

\section{Introduction}\label{introduction}

Generative artificial intelligence can assume part of the cognitive
effort required to draft, program, or formulate analyses. Experimental
evidence shows that it can reduce the cost of producing these outputs,
although the magnitude of its effects varies across tasks, users, and
organizational settings (Noy and Zhang 2023; Brynjolfsson, Li, and
Raymond 2025; Dell'Acqua et al. 2026). When such a reduction occurs, it
opens a cognitive margin that can be reallocated: users may produce
more, reduce their total effort, devote it to other activities, or
reinvest it in understanding, directing, and evaluating the assisted
task.

Reinvestment in the task is especially important because cognitive
savings do not by themselves guarantee a more reliable result.
Dell'Acqua et al. (2026) find substantial gains when tasks remain within
the frontier of AI capabilities, together with deteriorating performance
beyond it. Complementarily, Marcoccia, Quattrociocchi, and Capraro
(2026) show that receiving erroneous AI advice can reduce people's
willingness to suspend judgment and increase confidence in incorrect
answers. Taken together, these findings show that productivity gains can
coexist with conditions that make human evaluation more demanding.
Reduced effort is therefore the point of departure. The open question
concerns the destination of that margin and, in particular, the
conditions under which it is reinvested in preserving direction over the
process and its outcome.

The framework proposed by Fábrega (2026) (preprint) argues that this
reallocation should respond to the consequences users face when
delegating part of their cognitive effort. When a deficient output can
produce material, professional, or reputational costs, the expected
value of structuring the task, supervising its development, and
evaluating its outcome increases. Work-related activities often embed
generated products in performance and accountability relationships that
make those consequences more visible. Under comparable conditions, a
stronger work-related orientation should therefore be associated with a
greater willingness to reinvest effort in preserving human direction.
This expectation is probabilistic and allows for heterogeneity, because
different uses may also involve important consequences.

For this argument, human direction comprises the decisions through which
users structure, delimit, supervise, or correct the AI's contribution.
We distinguish two routes according to when that direction leaves
observable traces. Under \emph{specified delegation}, users organize the
task before or at the beginning of its execution through instructions,
code, outlines, constraints, tests, or validation rules. Execution can
then be delegated after the criteria guiding and constraining its
development have been defined. Under \emph{iterative coproduction},
direction is exercised over provisional outputs while the task unfolds.
Users may add context, question assumptions, request alternatives,
correct errors, or redirect the product through successive exchanges.

These routes represent different moments in the same effort to preserve
cognitive direction. They may coexist within a process and do not
constitute ordered levels of control. The first makes prior structuring
visible; the second makes intervention during production visible (Wu,
Terry, and Cai 2022; Beurer-Kellner, Fischer, and Vechev 2023; Huang et
al.~2026). An observable implication follows from this mechanism: if
work-related orientation raises the value of preserving direction,
activities with a greater presence of work-related uses should exhibit
stronger signals of specified delegation and iterative coproduction. The
possibility of observing these signals will, in turn, depend on the
window provided by each mode of use.

The \emph{Anthropic Economic Index} (AEI) makes it possible to examine
this implication because it distinguishes the work-related, personal,
and coursework composition of activities and separates interactions
conducted through Anthropic's first-party API (\emph{1P API}) from those
conducted through the Claude.ai conversational interface. Both modes may
contain the two routes of direction, although they provide different
observational windows. In Claude.ai, the data may preserve multi-turn
trajectories in which users clarify, correct, or redirect the course of
the interaction. In the \emph{1P API}, requests included in the index
are analyzed as isolated input-output pairs because calls lack metadata
that would link them to earlier exchanges. Part of the specification,
orchestration, validation, or iteration may therefore occur outside the
observed unit (Appel et al. 2026). This asymmetry affects the visibility
of traces without conceptually assigning a particular route of direction
to either mode.

This leads to the central question: to what extent is a stronger
work-related orientation, relative to personal and coursework
orientations, associated with aggregate signals of specified delegation
and iterative coproduction within the \emph{1P API} and \emph{Claude.ai}
modes? The analysis examines these associations separately within each
mode and at the level of aggregate AEI records. Each record corresponds
to a unique combination of mode of use, period, geography, activity
family, hierarchical level, and node, and contains metrics aggregated
for that combination. At this level, the data allow us to establish
whether the observed regularities are consistent with the implication
derived from the mechanism and how its expression varies across
observational windows. These records do not represent individuals,
organizations, or complete governance processes, nor do they directly
reveal who set the purpose, understood the outcome, or reallocated the
saved effort. Causal identification of that reallocation is therefore
beyond the scope of this study's design. The following sections
formalize the argument and present the empirical strategy used to test
its observable implications.

\section{Theoretical Framework and
Hypotheses}\label{theoretical-framework-and-hypotheses}

Research on agentic information systems proposes understanding their use
as a delegation relationship. Delegation involves transferring part of a
task, determining how much discretion the system receives, and
establishing how its capabilities are coordinated with human
capabilities (Baird and Maruping 2021). The automation literature had
also shown that a technology may intervene at different stages of an
activity without this alone determining who retains authority over the
process (Parasuraman, Sheridan, and Wickens 2000). This distinction
becomes especially relevant for generative AI. The autonomy and
inscrutability of these systems raise coordination and control problems
(Berente et al. 2021), while their probabilistic operation produces
outputs from patterns encoded during training. Riemer and Peter (2024)
characterize them as \emph{style engines} to highlight this generative
capacity and distinguish it from expectations of accuracy and
reliability inherited from conventional computing. When these products
take the form of texts, code, or analyses that we associate with
reasoning, delegating their execution may also shift part of the
discretion with which the task is developed (Shrestha, Ben-Menahem, and
Krogh 2019).

However, the fact that AI performs a considerable share of the
observable work does not necessarily mean that it controls the process.
A system may execute a task with high autonomy under conditions
carefully defined by people. It may also produce a similar amount of
work from an open-ended request, without prior criteria delimiting its
course. In both cases, the AI performs substantive operations. The
difference lies in who sets the purpose, establishes what counts as an
adequate outcome, and changes course when the product departs from those
criteria. The relevant question therefore goes beyond the distribution
of work volume between the person and the AI. It also requires
establishing where direction of the task remains.

Offloading cognitive operations frees attention and effort resources
(Risko and Gilbert 2016). People may reinvest those resources in other
components of the same task, for example by specifying the problem,
evaluating the result, or integrating its parts. They may also devote
them to a different activity, including leisure. Automation thus
substitutes for particular human operations and, at the same time, may
create new needs for formulation, evaluation, and coordination (Raisch
and Krakowski 2021; Jarrahi 2018). Effort savings do not determine which
of these trajectories will prevail.

We can understand the reallocation decision as a problem of cognitive
governance (Fábrega 2026). When incorporating generative AI, people
reorganize their effort between the operations they perform directly and
those they delegate. If they have incentives to retain epistemic
ownership of the task, they must reinvest part of the released effort in
directing its development, evaluating the product, integrating its
components, and answering for the final result. Here, epistemic
ownership refers to retaining sufficient understanding of the criteria
and reasons supporting the product so that the person can justify it and
revise its course. When those incentives are weak, the saved effort may
be reallocated away from the task, reducing human intervention in the
process. Governing the cognitive process thus consists of establishing
how these functions are distributed between people and AI and what
limits accompany delegation (Fábrega 2026).

This mechanism implies a relationship between the consequences of the
product and the value of preserving human direction. Loss of control is
more costly when an AI-mediated product informs a decision, must meet
recognizable standards, or is subject to third-party scrutiny. These
conditions occur more often in uses linked to productive processes.
Work-related outputs commonly circulate within organizations, affect
decisions, or generate professional and reputational consequences for
those who use them. In such contexts, people have stronger incentives to
retain criteria that guide execution and subsequently allow the outcome
to be evaluated. Of course, the argument operates in aggregate terms.
Some work-related uses have minor consequences, whereas certain personal
queries may be highly consequential for their users. Work-related
orientation serves as an approximation to contexts in which, on average,
the cost of insufficient direction increases. Accordingly, when the
composition of AI use shifts from personal to work-related purposes,
greater investment in observable forms of human direction should be
expected.

That direction may be exercised before execution or during its
development. The first route is \emph{specified delegation}. Here,
people structure the task in advance through instructions, code,
outlines, constraints, tests, or reusable procedures that delimit the
AI's action space (Wu, Terry, and Cai 2022; Beurer-Kellner, Fischer, and
Vechev 2023). An apparently precise instruction may still leave
substantive decisions open. Moreover, adequately representing the
purpose through instructions requires competencies that are unevenly
distributed among users (Zamfirescu-Pereira et al. 2023). Specification
constitutes a form of direction when the criteria introduced by the
person substantively define the task within which the AI will perform
its operations.

The second route is \emph{iterative coproduction}. In this
configuration, the AI generates a provisional output and the person
intervenes during its development to add context, question assumptions,
request alternatives, correct errors, or reformulate the chosen course.
The result is constructed through successive contributions from both
parties. The mere existence of several exchanges provides an
insufficient signal of direction. A conversation may be extended through
superficial modifications or new requests that leave the product's
orientation unchanged. Iteration constitutes direction when human
interventions substantively correct, edit, or refine the result or the
path used to obtain it (Amershi et al. 2019).

Both routes may coexist within the same process. A task structured in
advance may include subsequent revisions, while a conversational
trajectory may begin with a rigorous formulation of the problem. The
distinction identifies the principal moment when human direction leaves
observable traces. Under specified delegation, these traces concentrate
in the prior delimitation of the task. Under iterative coproduction,
they appear in interventions made over provisional outputs. This
temporal decomposition is consistent with recent evidence showing that
the mechanisms of collaboration between people and generative AI vary
across task phases (Huang et al. 2026). The phases studied by those
authors are not equivalent to the two routes proposed here, but they
support the analytical relevance of when the person intervenes.

The mode of use changes the conditions under which these interventions
can be performed and observed. The Anthropic Economic Index
distinguishes between Anthropic's first-party API (1P API) and the
Claude.ai conversational interface. Specified delegation and iterative
coproduction may exist in both environments. A conversation may begin
with detailed instructions, and an application connected through the API
may incorporate cycles of revision, validation, and reformulation. The
mode of use affects both the resources required to sustain those actions
and the traces available in the data.

Claude.ai organizes interaction as a conversational trajectory. Within
it, users can examine a provisional output and add information, question
an assumption, or request a new version without externally
reconstructing the task's state. Each intervention remains linked to
those that preceded it within the same exchange. This continuity reduces
the immediate cost of correcting or redirecting the product as it
develops and makes the sequence of interventions observable.

The API may also form part of iterative processes. An application may
chain calls, incorporate automated or human evaluations, and use their
results to modify later requests. However, the units published by the
AEI are analyzed as isolated input-output pairs because the calls do not
contain metadata that would link them to earlier exchanges. Part of the
orchestration, validation, or iteration may occur outside the observed
unit (Appel et al. 2026). The difference between the two modes therefore
concerns both the cost of sustaining successive intervention within the
interface and the possibility of observing it as part of the same
trajectory.

Figure 1 summarizes the two routes of direction and their relationship
to the observational environments.

\begin{figure}[H]

\centering{

\includegraphics[width=0.92\linewidth,height=\textheight,keepaspectratio]{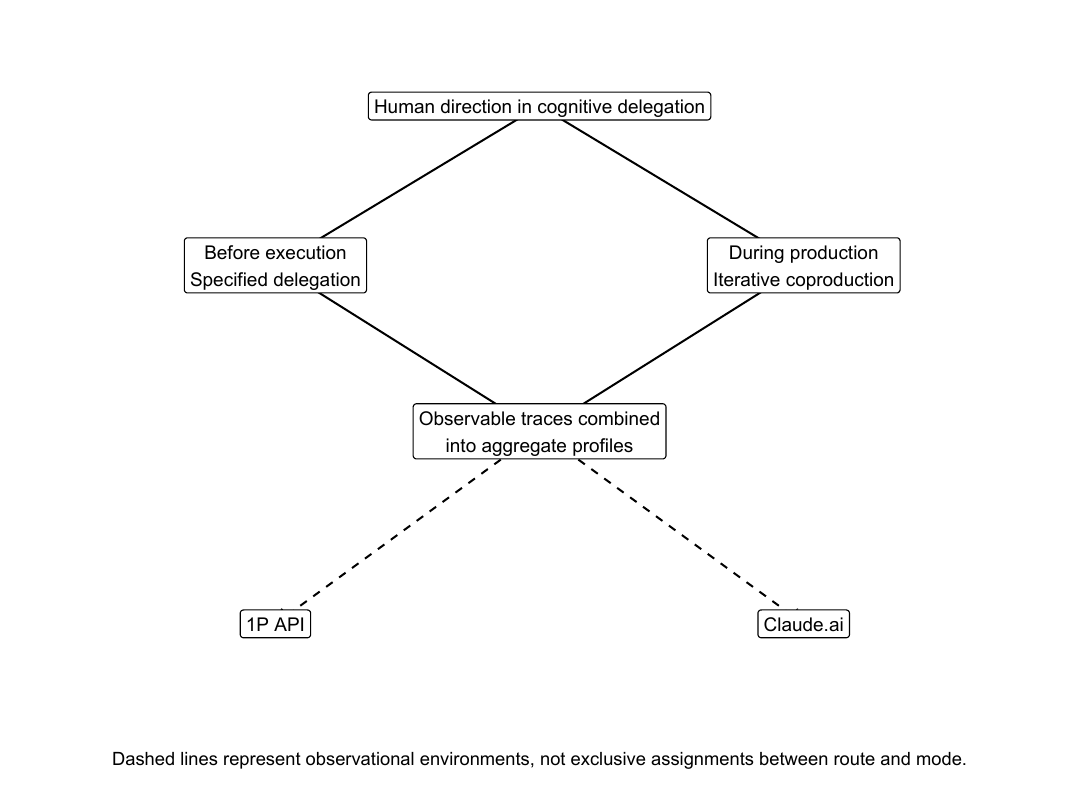}

}

\caption{\label{fig-conceptual}Two routes of direction and two
observational environments.}

\end{figure}%

The preceding mechanism supports a first expectation. As use shifts from
personal to work-related purposes, the expected adverse consequences of
insufficient task delimitation increase. People should therefore have
stronger incentives to direct human-AI interaction, for example by
setting criteria, constraints, or procedures that guide execution.
Because this form of direction can be incorporated into both 1P API and
Claude.ai interactions, the prediction concerns the change observed
within each mode:

\begin{quote}
\textbf{H1. A reallocation of use from personal purposes toward
work-related purposes will be associated with a more intense specified
delegation profile within each mode of use.}
\end{quote}

The same shift should also increase the incentive to intervene in
provisional outputs. The observable expression of that incentive depends
on the conditions of interaction. Claude.ai allows corrections and
redirections to be incorporated into a continuous conversational
trajectory. In the 1P API, iteration may require external orchestration
and remain outside the published unit. The increase in the iterative
profile associated with work-related orientation should therefore
manifest more strongly in Claude.ai:

\begin{quote}
\textbf{H2. The increase in the iterative coproduction profile
associated with this reallocation will be greater in Claude.ai than in
the 1P API.}
\end{quote}

H1 compares the change within each environment and is agnostic about
which mode exhibits the larger increase in specified delegation. H2
compares the magnitude of the iterative change across the two modes and
requires evaluating them over the same eligible activities and periods.
Both hypotheses allow the two routes of direction to coexist in 1P API
and Claude.ai.

\section{Data and Measurement}\label{data-and-measurement}

\subsection{Source, Panel Structure, and Unit of
Analysis}\label{source-panel-structure-and-unit-of-analysis}

To test the hypotheses, we use the public version of the \emph{Anthropic
Economic Index} (AEI) released on June 26, 2026. This release contains
metrics aggregated monthly for April and May 2026 and distinguishes two
activity sources. \texttt{source\_id\ =\ "1p\_api"} identifies direct
traffic through Anthropic's programming interface.
\texttt{source\_id\ =\ "claude\_ai"} combines Claude.ai activity,
including Chat and Cowork, from Free, Pro, and Max accounts. Claude Code
is excluded from both sources published in this release (Massenkoff et
al. 2026; Anthropic 2026). Hereafter, we refer to these modes as 1P API
and Claude.ai.

The original files organize the information in long format. Each row
associates an activity key with a metric (\texttt{metric\_id}) and its
value (\texttt{value}), so the same key appears repeatedly when
Anthropic publishes multiple metrics for it. For the analysis, we
transform these files to wide format by spreading the metrics into
columns. This operation changes the arrangement of the information
without aggregating, averaging, or altering the published values.

Each resulting observation is a monthly cell defined by nine key
variables, grouped into six dimensions: source (\texttt{source\_id});
the start and end dates of the period (\texttt{date\_start} and
\texttt{date\_end}); geographic identifier and level (\texttt{geo\_id}
and \texttt{geo\_level}); the family used to classify the activity
(\texttt{category\_name}); its hierarchical level
(\texttt{hierarchy\_level}); and the node's name and external identifier
(\texttt{node\_name} and \texttt{node\_external\_id}). The global panel
retains 20,532 monthly cells: 9,645 from 1P API and 10,887 from
Claude.ai. These cells represent aggregate activity categories. They do
not correspond to people, conversations, organizations, or complete work
sequences.

The AEI publishes alternative representations of part of the same
traffic through O*NET tasks, requests, SOC occupations, and overall
categories. It also presents these classifications at different
hierarchical levels. The families and their levels therefore do not
constitute independent samples that can be added to increase the number
of observations. The same underlying activity may be represented in more
than one view and at different degrees of aggregation.

To avoid this overlap, the main analysis uses a single view: level-0
O*NET tasks observed globally. In the structure published by the AEI,
this is the most granular level of the O*NET classification. The main
view contains 4,285 1P API cells and 5,111 Claude.ai cells. The
remaining families and levels are reserved for examining the scope of
the findings. The published values, transformation rules, and
construction of the different views are documented in Appendix S1.

\subsection{Construction of the Direction
Profiles}\label{construction-of-the-direction-profiles}

The human direction proposed in the theoretical framework is not
directly recorded in the AEI. The available metrics describe aggregate
classifications of observed behavior and estimated characteristics of
tasks, inputs, and responses. Each provides a partial trace of human-AI
interaction. Combining them makes it possible to construct profiles
consistent with specified delegation and iterative coproduction,
although it does not allow us to observe the complete mechanism or
attribute it to a particular person.

We use six metrics. Two describe how collaboration is organized.
Directive execution, \(D_g\), is the share of activity classified as
\emph{directive execution}, in which a person formulates a task for
Claude to execute. Iteration, \(I_g\), records the share classified as
\emph{task iteration}, in which the person successively refines a
product. Because each cell combines multiple interactions, both forms
may coexist in different proportions within the same aggregate
observation.

The remaining four metrics situate those patterns in context. Autonomy,
\(A_g\), records the mean degree of autonomy granted to Claude. The
variable \(L_g^H\) represents the relative position of the estimated
reading demand required to understand the human input, while
\(L_g^{AI}\) does the same for the generated response. These two
variables describe characteristics of the texts, not the user's actual
education, the correctness of the response, or the quality of the
product. Finally, \(H_g\) indicates the share of activity corresponding
to tasks that a competent human could complete without AI. This last
metric characterizes the task's potential accessibility to human
judgment without measuring the capabilities of the people who produced
the interactions.

Before combining these signals, we place them on a common scale from
zero to one. The shares of directive execution, iteration, and estimated
human capability are divided by one hundred. Autonomy, published on a
one-to-five scale, is transformed as \(A_g=(x_g-1)/4\). The estimated
reading demands are converted into cumulative positions within a common
reference distribution, constructed from level-0 O*NET tasks observed
globally in April 2026 and pooling both modes of access. This reference
remains fixed throughout the analysis, so that the same reading demand
occupies the same relative position when comparing sources, months, and
scope exercises.

The behavioral trace closest to specified delegation is \(D_g\), because
it identifies activity in which a person formulates a task for Claude to
execute. This signal is insufficient by itself: a request may be
directive while leaving substantive decisions open. We therefore combine
it with three conditions that approximate how extensively execution was
delimited in advance. \(L_g^H\) provides a signal of the relative
elaboration of the human input; \(H_g\) indicates that the task remains
potentially accessible to a competent human; and \((1-A_g)\) increases
when the discretion granted to the system is lower. The specified
delegation profile is defined as:

\begin{equation}\phantomsection\label{eq-de}{
SD_g=\left[D_g(1-A_g)L_g^H H_g\right]^{1/4}.
}\end{equation}

Iterative coproduction places direction during the development of the
product. Its point of departure is \(I_g\), which records the aggregate
presence of iterative refinement. To distinguish substantive
intervention from the mere repetition of exchanges, we incorporate the
relative elaboration of the human input, \(L_g^H\), and the reading
demand of the response under review, \(L_g^{AI}\). Under this route,
\(A_g\) enters positively because some autonomy allows the AI to produce
a provisional contribution that the person can correct, expand, or
redirect. The profile is defined as:

\begin{equation}\phantomsection\label{eq-ci}{
IC_g=\left[I_gL_g^H L_g^{AI}A_g\right]^{1/4}.
}\end{equation}

Estimated human capability, \(H_g\), is included only in specified
delegation. Iterative direction may be exercised over a product that the
person could not have generated completely without assistance, provided
that the person can evaluate and transform it. In turn, \(L_g^{AI}\)
appears only in coproduction because this route is exercised over a
product generated during the interaction. These differences carry into
measurement the distinct moment when human direction becomes observable.

Both profiles use geometric means because their components operate as
formative signals that must concur. We do not assume that they are
interchangeable manifestations of a single latent trait (Petter, Straub,
and Rai 2007). A sum would allow a high value to compensate completely
for the absence of another condition. Multiplication limits such
compensation: intensely directive activity provides little evidence of
specified delegation when the other signals of prior delimitation are
weak, just as high reading demand provides little evidence of
coproduction when no iteration is observed. The fourth root returns the
product to the original scale from zero to one (OECD, European Union,
and European Commission, Joint Research Centre 2008). For ease of
presentation, we report the profiles in points on a zero-to-one-hundred
scale.

Equations 1 and 2 do not include the work-related composition of the
activity. The association between that composition and the profiles is
therefore an empirical finding rather than a relationship introduced
through their definitions. Interpretation remains bounded at the level
of published cells. A high value identifies an aggregate configuration
in which signals consistent with one of the theoretical routes concur;
it does not demonstrate that a specific person defined the purpose,
reviewed the reasoning, or understood the reasons supporting the
outcome. Analyses that decompose the indicators into a core and a
condition, together with other auxiliary equations, are presented in
Appendix S2.

\section{Empirical Strategy}\label{empirical-strategy}

\subsection{Use-Case Recomposition and
Estimands}\label{use-case-recomposition-and-estimands}

The hypotheses link the direction profiles to the composition of the
activity's assigned purpose. The AEI reports three shares: work-related
use (\texttt{use\_case\_work\_pct}), personal use
(\texttt{use\_case\_personal\_pct}), and \texttt{coursework}
(\texttt{use\_case\_coursework\_pct}). After dividing the percentages by
one hundred, we retain and normalize compositions whose published sum
lies between 0.95 and 1.05. We denote the three parts by \(w_g\),
\(p_g\), and \(c_g\), respectively, such that:

\[
w_g+p_g+c_g=1.
\]

The \texttt{coursework} component remains the third technical part of
the composition and the omitted category in the models. The current
hypotheses do not require a substantive interpretation of its distance
from the other uses. The contrast holds this share constant and shifts
ten percentage points from the personal component to the work-related
component:

\begin{equation}\phantomsection\label{eq-shift}{
(w_g,p_g,c_g)\longrightarrow(w_g+0.10,p_g-0.10,c_g).
}\end{equation}

This recomposition compares two predictions for the same cell: one under
its observed composition and another under a composition that is ten
points more work-related and ten points less personal. Source, month,
and activity node remain constant. The contrast indicates how much the
expected profile would vary if the same monthly category had a
relatively stronger orientation toward work-related use. Because the AEI
contains aggregate observational data, we interpret this difference as
an adjusted association between the composition of uses and the profile.
It does not represent the causal effect of changing the motivation of an
individual interaction.

H1 is evaluated separately within 1P API and Claude.ai. For each source,
we estimate the average expected change in \(SD_g\) between the observed
composition and the recomposition defined in Equation~\ref{eq-shift}.
The contrast is calculated over cells with \(p_g\geq0.10\), because only
those cells permit shifting ten points from the personal component.

H2 compares the response of the iterative profile across the two modes.
For each eligible cell, we calculate the expected change in \(IC_g\) and
then estimate the difference between the average change in Claude.ai and
that in 1P API. This comparison uses exclusively nodes and months that
are present in both sources and eligible for the same recomposition. The
hypothesis is therefore evaluated over common activities and periods
rather than confounding the between-mode difference with their different
task compositions.

\subsection{Models, Support, and
Uncertainty}\label{models-support-and-uncertainty}

The profiles are bounded between zero and one. The main specification
therefore uses quasibinomial fractional-response models with a logit
link (Papke and Wooldridge 1996). For H1, we estimate separate models by
source with node and month fixed effects. These models relate the
work-related and personal composition to \(SD_g\) within each
environment and allow the association to take a different form in 1P API
and Claude.ai without imposing an ordering between the two modes.

For H2, we first restrict the data to the common support. On this set,
we fit a model of \(IC_g\) that includes interactions between use
composition and source:

\begin{equation}\phantomsection\label{eq-model}{
\operatorname{logit}\{E(Y_g\mid X_g)\}
=
\alpha+\beta_w w_g+\beta_p p_g+
r_g(\delta_0+\delta_w w_g+\delta_p p_g)
+\mu_{n(g)}+\lambda_{t(g)}.
}\end{equation}

In Equation~\ref{eq-model}, \(r_g\) identifies Claude.ai observations,
\(\mu_{n(g)}\) represents node fixed effects, and \(\lambda_{t(g)}\)
represents month effects. \texttt{coursework} remains the omitted part
because of the compositional constraint. For H1, we use the
corresponding within-source version of the model without the \(r_g\)
indicator or its interactions.

The model coefficients depend on the omitted compositional category and
are expressed on the logit scale, so they are not the substantive object
of interpretation. We evaluate the hypotheses using standardized
predictions. For each eligible cell, we obtain the expected profile
under the two compositions in Equation~\ref{eq-shift}, calculate their
difference, and then average these changes over the relevant support.
This procedure preserves the observed distribution of nodes and months
and translates the models into changes expressed in profile points.

The strict support used for H2 contains 1,411 node-month pairs and 870
nodes. The restriction holds the published activity category and period
constant, but it does not equate the people, organizations, or processes
that generated traffic in each environment. The comparison concerns the
average change in profiles constructed consistently from the telemetry
available in each mode.

Nodes may appear in both months and, in the H2 contrast, in both
sources. This structure induces dependence among observations belonging
to the same activity category. We estimate uncertainty through a cluster
bootstrap that resamples nodes with replacement while jointly retaining
all their months. For H1, resampling occurs within each source. For H2,
we jointly resample common nodes so that each pair and its two sources
receive the same replication weight.

In each of 2,000 successful replications, we reconstruct the
corresponding sample, reestimate the models, and recalculate the changes
associated with the hypotheses and the between-mode difference. This
strategy incorporates the covariance between 1P API and Claude.ai in H2
and avoids inferring the difference by comparing marginal intervals. We
report 95\% percentile intervals.

\section{Results}\label{results}

\subsection{Specified Delegation within Each
Mode}\label{specified-delegation-within-each-mode}

A ten-point reallocation from personal to work-related use is associated
with an increase of 2.76 points in the specified delegation profile for
1P API ({[}2.30, 3.22{]}) and 1.45 points for Claude.ai ({[}0.93,
1.97{]}). Both intervals remain above zero (see Figure 2 and Table 1).
The evidence is consistent with H1 within each mode.

\begin{figure}[H]

\centering{

\includegraphics[width=0.93\linewidth,height=\textheight,keepaspectratio]{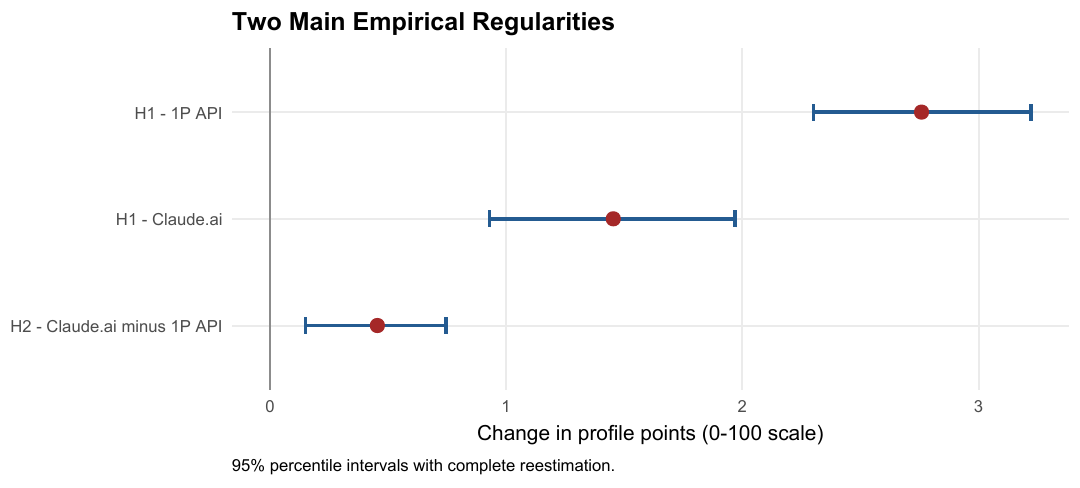}

}

\caption{\label{fig-main}Main results for H1 and H2.}

\end{figure}%

\begin{longtable}[]{@{}
  >{\raggedright\arraybackslash}p{(\linewidth - 6\tabcolsep) * \real{0.2500}}
  >{\raggedright\arraybackslash}p{(\linewidth - 6\tabcolsep) * \real{0.2500}}
  >{\raggedright\arraybackslash}p{(\linewidth - 6\tabcolsep) * \real{0.2500}}
  >{\raggedright\arraybackslash}p{(\linewidth - 6\tabcolsep) * \real{0.2500}}@{}}
\caption{Two hypotheses and their main estimands. Each interval uses
2,000 successful replications.}\tabularnewline
\toprule\noalign{}
\begin{minipage}[b]{\linewidth}\raggedright
Hyp.
\end{minipage} & \begin{minipage}[b]{\linewidth}\raggedright
Estimand
\end{minipage} & \begin{minipage}[b]{\linewidth}\raggedright
Change {[}95\% CI{]}
\end{minipage} & \begin{minipage}[b]{\linewidth}\raggedright
Interpretation
\end{minipage} \\
\midrule\noalign{}
\endfirsthead
\toprule\noalign{}
\begin{minipage}[b]{\linewidth}\raggedright
Hyp.
\end{minipage} & \begin{minipage}[b]{\linewidth}\raggedright
Estimand
\end{minipage} & \begin{minipage}[b]{\linewidth}\raggedright
Change {[}95\% CI{]}
\end{minipage} & \begin{minipage}[b]{\linewidth}\raggedright
Interpretation
\end{minipage} \\
\midrule\noalign{}
\endhead
\bottomrule\noalign{}
\endlastfoot
H1 & 1P API & 2.76 {[}2.30, 3.22{]} & Positive change within the mode \\
H1 & Claude.ai & 1.45 {[}0.93, 1.97{]} & Positive change within the
mode \\
H2 & Claude.ai minus 1P API & 0.45 {[}0.15, 0.75{]} & Positive
difference on the main support \\
\end{longtable}

\subsection{Iterative Coproduction in Common
Activities}\label{iterative-coproduction-in-common-activities}

Across the same eligible O*NET tasks and months, the iterative profile
changes by -0.30 points in 1P API ({[}-0.75, 0.15{]}) and by 0.15 in
Claude.ai ({[}-0.30, 0.60{]}). The Claude.ai response is positive,
whereas the 1P API response is negative in this restricted specification
(see Figure 3).

The Claude.ai-minus-1P API difference reaches 0.45 points ({[}0.15,
0.75{]}). Because it is calculated within each replication, its interval
incorporates covariance between sources. This pattern is consistent with
H2: iterative coproduction is more strongly related to work-related
orientation in Claude.ai than in 1P API on the main common support.

\begin{figure}[H]

\centering{

\includegraphics[width=0.93\linewidth,height=\textheight,keepaspectratio]{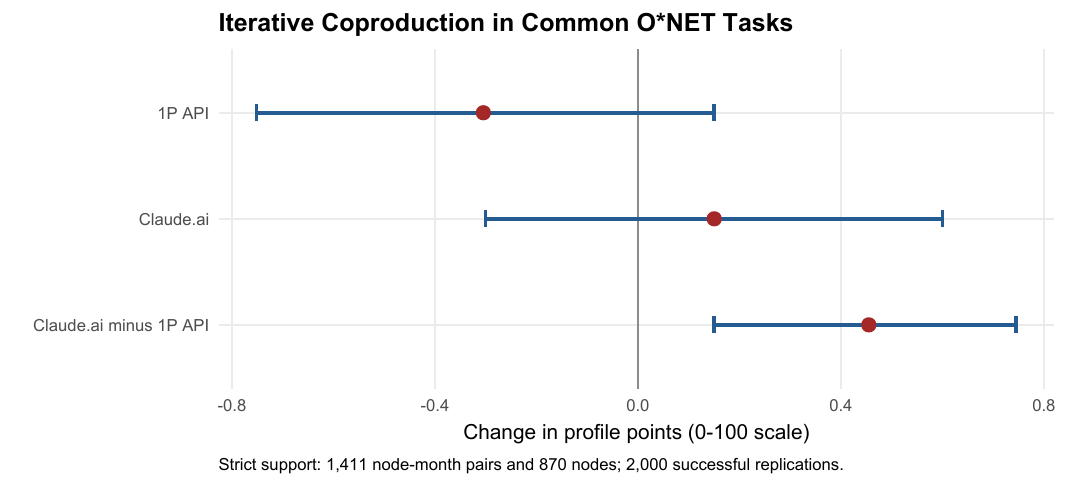}

}

\caption{\label{fig-common}Iterative coproduction in common O*NET
tasks.}

\end{figure}%

\subsection{Scope and Sensitive
Findings}\label{scope-and-sensitive-findings}

The iterative difference remains positive when the analysis is repeated
separately at other O*NET levels and for requests and SOC occupations.
These taxonomies represent traffic from alternative angles and are not
added together. The complete results, together with the corresponding
pairs and nodes, appear in Appendix S4.

Alternative aggregators and ILR coordinates further modify the
magnitudes and, in some specifications, the sign of the 1P API result.
By contrast, Claude.ai retains a positive iterative response across the
aggregations examined. The geographic extension within Claude.ai is also
positive (Appendices S5 and S6).

\section{Conclusions}\label{conclusions}

The evidence presented shifts the analysis from how much work the AI
executes toward how direction of the task is organized. A stronger
work-related orientation is associated with a more intense specified
delegation profile in both 1P API and Claude.ai. In addition, across
common O*NET tasks and periods, the increase in the iterative
coproduction profile is greater in Claude.ai. These findings identify
two ways in which human direction may leave traces: before execution,
through criteria that delimit the task, and during production, through
interventions in provisional outputs.

This distinction extends discussions of automation, augmentation, and
delegation. These literatures examine how execution, autonomy, and
discretion are distributed between people and systems (Raisch and
Krakowski 2021; Baird and Maruping 2021). Our argument incorporates when
direction is exercised. This temporal dimension speaks to evidence
showing that human-AI collaboration activates different mechanisms
across task phases (Huang et al. 2026). A similar amount of delegated
execution may be part of processes organized through prior
specifications or of others that retain human intervention during
production.

The iterative difference between modes requires careful interpretation.
Claude.ai preserves a conversational trajectory in which revisions can
be linked to the products that prompted them. 1P API calls are observed
as isolated input-output pairs, although part of their orchestration,
validation, or iteration may occur externally. The stronger relationship
observed in Claude.ai may reflect both the intervention possibilities
afforded by a conversational trajectory and the greater visibility of
those interventions. The available data do not allow us to separate
these two components. The result therefore describes a difference in the
observable profile and leaves open the complete organization of
direction in each environment.

By reducing the cost of producing cognitive outputs, generative AI gives
humans incentives to redistribute their cognitive effort toward
functions of formulation, evaluation, integration, and accountability
(Fábrega 2026). Someone must translate purposes and domain knowledge
into execution conditions, evaluate the product, and incorporate it into
a decision. Limited human intervention within the observed unit may
coexist with considerable direction work performed before or outside it.
This perspective suggests organizational implications. Stable and
repetitive tasks may rely on specifications, tests, and reusable rules.
Ambiguous or context-sensitive tasks, however, require opportunities to
examine provisional outputs and correct their course. Processes with
substantial consequences will probably require both forms. In such
processes, it matters who can modify specifications, what criteria
govern acceptance of the product, and when the process must stop or be
escalated. These possibilities facilitate direction, although they do
not guarantee that the person understands the outcome or can justify it.

The empirical scope of this study remains limited. The profiles combine
behavioral classifications with estimated characteristics of tasks and
texts. Aggregate cells do not identify users, incentives, actual
understanding, or organizational rules. The common support holds
published activities and periods constant but does not equate the people
or organizations that generated the traffic. The findings express
associations, depend on two months of activity from a single provider,
and may reflect classification error and differences in observability
across modes.

These limitations define a research agenda. Data are needed that
reconstruct complete processes by linking prior specifications,
interaction sequences, revisions, acceptance decisions, and subsequent
consequences. For the API, this requires observing the external systems
that chain and validate calls. Longitudinal designs and field
experiments could evaluate how task consequences and oversight
mechanisms affect error detection, the ability to provide justification,
and subsequent unaided performance. This program would make it possible
to compare people and organizations according to how they distribute
direction and accountability across their human-AI cognitive processes.

\enlargethispage{\baselineskip}

\section*{Data Availability}\label{data-availability}
\addcontentsline{toc}{section}{Data Availability}

The analysis uses the public version of the \emph{Anthropic Economic
Index} released on June 26, 2026. Reproducibility will be provided later
on GitHub.

\break

\section*{Appendix S1. Sources, Panel Construction, and
Traceability}\label{appendix-s1}
\addcontentsline{toc}{section}{Appendix S1. Sources, Panel Construction,
and Traceability}

The public release of the \emph{Anthropic Economic Index} dated June 26,
2026, contains two files. \texttt{aei\_1p\_api\_2026-06-26.csv}
represents direct calls to Anthropic's 1P API and excludes Claude Code.
\texttt{aei\_claude\_ai\_2026-06-26.csv} represents Chat and Cowork
activity from Free, Pro, and Max accounts; the article refers to this
source as Claude.ai (Massenkoff et al. 2026; Anthropic 2026).

The files contain 491,705 rows for 1P API and 1,636,573 for Claude.ai.
Each row reports one metric for a published combination of geography,
family, level, and node. Processing validates the keys and converts
\texttt{metric\_id} values into columns without aggregating or averaging
values. The key for a global cell combines source, start and end dates,
geography, classification family, hierarchical level, and the node's
external identifier.

The wide panel contains 20,532 global monthly cells: 9,645 from 1P API
and 10,887 from Claude.ai. There are ten representations: four O*NET
levels, three request levels, two SOC occupational levels, and one
overall view. Level 0 of O*NET corresponds to tasks and is the most
granular level. The taxonomies and levels are alternative
representations of part of the same traffic, not independent units.

The shares \texttt{use\_case\_work\_pct},
\texttt{use\_case\_personal\_pct}, and
\texttt{use\_case\_coursework\_pct} are divided by one hundred. When
their sum lies between 0.95 and 1.05, each part is divided by that sum.
The procedure produces 20,531 complete compositions; one 1P API cell
remains incomplete. \texttt{coursework} is retained only to close the
composition and as the omitted part of the parameterization.

\begin{longtable}[]{@{}
  >{\raggedright\arraybackslash}p{(\linewidth - 14\tabcolsep) * \real{0.2045}}
  >{\raggedright\arraybackslash}p{(\linewidth - 14\tabcolsep) * \real{0.0795}}
  >{\raggedleft\arraybackslash}p{(\linewidth - 14\tabcolsep) * \real{0.0682}}
  >{\raggedleft\arraybackslash}p{(\linewidth - 14\tabcolsep) * \real{0.1818}}
  >{\raggedleft\arraybackslash}p{(\linewidth - 14\tabcolsep) * \real{0.1705}}
  >{\raggedleft\arraybackslash}p{(\linewidth - 14\tabcolsep) * \real{0.1477}}
  >{\raggedleft\arraybackslash}p{(\linewidth - 14\tabcolsep) * \real{0.0682}}
  >{\raggedleft\arraybackslash}p{(\linewidth - 14\tabcolsep) * \real{0.0795}}@{}}

\caption{\label{tbl-s1-support}Coverage of the main view and strict
support}

\tabularnewline

\toprule\noalign{}
\begin{minipage}[b]{\linewidth}\raggedright
Analysis
\end{minipage} & \begin{minipage}[b]{\linewidth}\raggedright
Family
\end{minipage} & \begin{minipage}[b]{\linewidth}\raggedleft
Level
\end{minipage} & \begin{minipage}[b]{\linewidth}\raggedleft
Estimation rows
\end{minipage} & \begin{minipage}[b]{\linewidth}\raggedleft
Eligible cells
\end{minipage} & \begin{minipage}[b]{\linewidth}\raggedleft
Common pairs
\end{minipage} & \begin{minipage}[b]{\linewidth}\raggedleft
Nodes
\end{minipage} & \begin{minipage}[b]{\linewidth}\raggedleft
Months
\end{minipage} \\
\midrule\noalign{}
\endhead
\bottomrule\noalign{}
\endlastfoot
H1 1P API & onet & 0 & 4285 & 1536 & & 2427 & 2 \\
H1 Claude.ai & onet & 0 & 5111 & 3425 & & 2815 & 2 \\
H2 common support & onet & 0 & 2822 & 2822 & 1411 & 870 & 2 \\

\end{longtable}

SHA-256 hashes of the original data are verified before and after the
pipeline. The derived data retain missing values as missing and generate
a canonical node identifier that prevents collisions across families and
levels. The R session and package information is recorded in the
replication package.

\break

\section*{Appendix S2. Construction and Decomposition of the
Profiles}\label{appendix-s2}
\addcontentsline{toc}{section}{Appendix S2. Construction and
Decomposition of the Profiles}

The article retains the equations for the complete profiles. This
appendix documents their auxiliary layers. For specified delegation, the
core combines directive execution and estimated human capability:

\[
K_g^{SD}=(D_gH_g)^{1/2}.
\]

The specification condition combines relative input elaboration and
restricted autonomy:

\[
E_g^{SD}=\left[L_g^H(1-A_g)\right]^{1/2}.
\]

By construction, \(SD_g=\sqrt{K_g^{SD}E_g^{SD}}\). The decomposition
adds no components and does not change the profile. It indicates whether
a difference is concentrated in directive behavior and the task's human
accessibility or in conditions consistent with prior delimitation.

For iterative coproduction, the core combines iteration with the
relative elaboration of the response:

\[
K_g^{IC}=(I_gL_g^{AI})^{1/2}.
\]

The condition combines input elaboration and positive autonomy:

\[
E_g^{IC}=(L_g^HA_g)^{1/2},
\qquad IC_g=\sqrt{K_g^{IC}E_g^{IC}}.
\]

Autonomy serves different functions in the two configurations. Lower
autonomy is consistent with prior delimitation; positive autonomy allows
a provisional output to exist so that a person can intervene in it. In
neither case does autonomy by itself determine human direction.

\subsection*{Reconstruction Example}\label{reconstruction-example}
\addcontentsline{toc}{subsection}{Reconstruction Example}

The global 1P API cell for May 2026 corresponding to the O*NET task
``Handling and Moving Objects,'' observed at level 3, illustrates the
rule. Directive execution is 0.7659, autonomy is 0.3225, the relative
position of the human input is 0.1740, and estimated human capability is
0.9366. Their combination produces \(SD_g=0.5393\). Iteration is 0.0293
and the relative position of the response is 0.1161; with the same input
and autonomy, \(IC_g=0.1175\).

The difference between the two values does not order the routes on a
general scale of direction. It indicates that, for this cell, the
constitutive traces of specified delegation concur more strongly.
\texttt{profile\_identity\_audit.csv} verifies that the reconstruction
of both profiles matches the analytical objects within a tolerance of
\(10^{-12}\).

\break

\section*{Appendix S3. Estimands, Support, and
Bootstrap}\label{appendix-s3}
\addcontentsline{toc}{section}{Appendix S3. Estimands, Support, and
Bootstrap}

The compositional intervention holds \texttt{coursework} constant and
shifts ten points from personal to work-related use. H1 uses separate
source models over level-0 O*NET tasks. The models are estimated with
all available complete cells, and changes are averaged only across cells
whose personal component permits the recomposition.

H2 requires node-month pairs that are present and eligible in both
sources. The same set is used to fit the model with node and month fixed
effects, predict both changes, and calculate Claude.ai minus 1P API. The
support contains 1,411 pairs and 870 nodes.

The bootstrap uses seed 20260810. For H1, it resamples nodes within each
source. For H2, it jointly resamples nodes from the common support; both
sources in each pair receive the same replication weight. Each
replication reconstructs the resampled set, reestimates the
quasibinomial models, and recalculates predictions. Execution continues
until 2,000 successful replications are obtained and uses 95\%
percentile intervals.

\begin{longtable}[]{@{}
  >{\raggedright\arraybackslash}p{(\linewidth - 10\tabcolsep) * \real{0.2872}}
  >{\raggedleft\arraybackslash}p{(\linewidth - 10\tabcolsep) * \real{0.0745}}
  >{\raggedleft\arraybackslash}p{(\linewidth - 10\tabcolsep) * \real{0.1383}}
  >{\raggedright\arraybackslash}p{(\linewidth - 10\tabcolsep) * \real{0.1489}}
  >{\raggedleft\arraybackslash}p{(\linewidth - 10\tabcolsep) * \real{0.1383}}
  >{\raggedleft\arraybackslash}p{(\linewidth - 10\tabcolsep) * \real{0.2128}}@{}}

\caption{\label{tbl-s3-bootstrap}Main estimates with complete
reestimation}

\tabularnewline

\toprule\noalign{}
\begin{minipage}[b]{\linewidth}\raggedright
Estimand
\end{minipage} & \begin{minipage}[b]{\linewidth}\raggedleft
Point
\end{minipage} & \begin{minipage}[b]{\linewidth}\raggedleft
Bootstrap SE
\end{minipage} & \begin{minipage}[b]{\linewidth}\raggedright
95\% CI
\end{minipage} & \begin{minipage}[b]{\linewidth}\raggedleft
Replications
\end{minipage} & \begin{minipage}[b]{\linewidth}\raggedleft
Positive proportion
\end{minipage} \\
\midrule\noalign{}
\endhead
\bottomrule\noalign{}
\endlastfoot
H1: 1P API & 2.758 & 0.231 & {[}2.30, 3.22{]} & 2000 & 1.000 \\
H1: Claude.ai & 1.454 & 0.265 & {[}0.93, 1.97{]} & 2000 & 1.000 \\
H2: 1P API change & -0.305 & 0.231 & {[}-0.75, 0.15{]} & 2000 & 0.100 \\
H2: Claude.ai change & 0.150 & 0.233 & {[}-0.30, 0.60{]} & 2000 &
0.732 \\
H2: Claude.ai minus 1P API & 0.455 & 0.152 & {[}0.15, 0.75{]} & 2000 &
0.998 \\

\end{longtable}

The interval for the difference is not formed by subtracting marginal
endpoints. The difference is calculated within each replication,
incorporating covariance between the two sources. Inference is
conditional on the published cells, the two available months, the AEI
classifications, and the support rules.

\break

\section*{Appendix S4. Common Activities and Scope across
Representations}\label{appendix-s4}
\addcontentsline{toc}{section}{Appendix S4. Common Activities and Scope
across Representations}

The main view uses level-0 O*NET tasks because it provides a granular,
non-overlapping representation. To evaluate scope, the same
strict-support procedure is repeated separately at other O*NET levels
and for requests and SOC occupations. Each row is a distinct estimate;
pairs and nodes must not be added across rows.

\begin{longtable}[]{@{}lrrrl@{}}

\caption{\label{tbl-s4-scope}Scope of the iterative contrast across
alternative representations}

\tabularnewline

\toprule\noalign{}
Family & Level & Common pairs & Nodes & Claude.ai minus 1P API \\
\midrule\noalign{}
\endhead
\bottomrule\noalign{}
\endlastfoot
O*NET & 0 & 1411 & 870 & 0.45 \\
O*NET & 1 & 509 & 298 & 0.72 \\
O*NET & 2 & 195 & 111 & 1.16 \\
O*NET & 3 & 44 & 23 & 0.51 \\
Requests & 0 & 833 & 473 & 0.87 \\
Requests & 1 & 214 & 112 & 1.29 \\
Requests & 2 & 24 & 12 & 0.76 \\
SOC occupations & 0 & 660 & 377 & 1.27 \\
SOC occupations & 1 & 27 & 15 & 0.36 \\

\end{longtable}

The difference remains positive across all listed representations.
Higher levels, however, group tasks and reduce the number of nodes. Sign
stability does not turn these rows into independent replications of the
same phenomenon; it shows that the main ordering does not depend
exclusively on a particular taxonomy.

\break

\section*{Appendix S5. Geometry and Definition of the
Measure}\label{appendix-s5}
\addcontentsline{toc}{section}{Appendix S5. Geometry and Definition of
the Measure}

Each cell receives equal weight. We do not use \texttt{pct} as a weight
because it is an aggregate prevalence by source, month, and node,
published after aggregation thresholds; it does not identify people or
their individual motivations and objectives. Moreover, \texttt{pct}
values differ between 1P API and Claude.ai. Applying them as weights
would give the same tasks different influence in each mode and would
combine the contrast within the common support with differences in the
aggregate composition of traffic. Equal weighting preserves a direct and
transparent comparison across published cells.

Isometric log-ratio coordinates modify the functional geometry of the
composition (Aitchison 1986; Egozcue et al. 2003). Arithmetic, harmonic,
and minimum aggregators modify the construct, not only the model form.

\begin{longtable}[t]{lllll}

\caption{\label{tbl-s5-sensitivity}Geometry and measurement
sensitivities}

\tabularnewline

\\
\toprule
Exercise & Profile & Mode or contrast & Points & N\\
\midrule
\endfirsthead
\caption[]{Geometry and measurement sensitivities (continued)}\\
\toprule
Exercise & Profile & Mode or contrast & Points & N\\
\midrule
\endhead

\endfoot
\bottomrule
\endlastfoot
ILR geometry & Specified delegation & 1P API & 3.03 & 4,491\\
ILR geometry & Specified delegation & Claude.ai & 3.68 & 8,657\\
ILR geometry & Iterative coproduction & 1P API & -0.62 & 4,491\\
ILR geometry & Iterative coproduction & Claude.ai & 4.15 & 8,657\\
Alternative aggregator: arithmetic mean & Iterative coproduction & 1P API & 3.64 & 4,491\\
\addlinespace
Alternative aggregator: arithmetic mean & Iterative coproduction & Claude.ai & 4.03 & 8,657\\
Alternative aggregator: harmonic mean & Iterative coproduction & 1P API & 0.21 & 4,491\\
Alternative aggregator: harmonic mean & Iterative coproduction & Claude.ai & 4.09 & 8,657\\
Alternative aggregator: minimum & Iterative coproduction & 1P API & -0.08 & 4,491\\
Alternative aggregator: minimum & Iterative coproduction & Claude.ai & 3.09 & 8,657\\
\addlinespace
Alternative aggregator: arithmetic mean & Specified delegation & 1P API & 2.82 & 4,491\\
Alternative aggregator: arithmetic mean & Specified delegation & Claude.ai & 1.52 & 8,657\\
Alternative aggregator: harmonic mean & Specified delegation & 1P API & 5.75 & 4,491\\
Alternative aggregator: harmonic mean & Specified delegation & Claude.ai & 4.36 & 8,657\\
Alternative aggregator: minimum & Specified delegation & 1P API & 6.48 & 4,491\\
\addlinespace
Alternative aggregator: minimum & Specified delegation & Claude.ai & 3.51 & 8,657\\
Nonlinear link & Specified delegation & 1P API & 5.67 & 4,491\\
Nonlinear link & Specified delegation & Claude.ai & 3.39 & 8,648\\
Nonlinear link & Iterative coproduction & 1P API & 1.89 & 4,491\\
Nonlinear link & Iterative coproduction & Claude.ai & 4.84 & 8,649\\*

\end{longtable}

Interpretation must distinguish two classes of change. Rotating the
omitted category preserves predictions algebraically. Changing to ILR
alters the functional relationship among parts; changing the aggregator
alters the definition of the profile. These specifications examine
geometry and measurement and do not replace the main estimand over
common level-0 O*NET tasks.

\break

\section*{Appendix S6. Geographic Extension within
Claude.ai}\label{appendix-s6}
\addcontentsline{toc}{section}{Appendix S6. Geographic Extension within
Claude.ai}

Geographic disaggregation is available only for Claude.ai and, within
the O*NET classification, only at hierarchy level 3. The geographic
analysis therefore uses a different analytical support from the main H2
specification, which relies on level-0 ONET tasks observed in both
Claude.ai and 1P API. The magnitudes reported here are consequently not
directly comparable with the main estimands. Within the level-3
Claude.ai data, the global iterative change is positive, country
estimates are also positive, and sequential country exclusion preserves
the pattern. This extension documents geographic persistence within
Claude.ai under the available level of aggregation. Explaining the
difference in magnitude relative to the main specification requires
further analysis and lies beyond the scope of the present study.

\begin{table}

\caption{\label{tbl-s6-geography}Geographic extension of iterative
coproduction in Claude.ai}

\centering{

[!h]
\centering\begingroup\fontsize{9}{11}\selectfont

\begin{tabular}{>{\raggedright\arraybackslash}p{5.0cm}r>{\raggedleft\arraybackslash}p{2.7cm}r>{\raggedleft\arraybackslash}p{2.4cm}}
\toprule
Summary & Estimate & Interval or range & Countries & Max. eligible observations per country\\
\midrule
Global estimate & 5.55 & {}[5.09, 5.97] & 76 & 3034\\
Countries: median and range & 5.83 & {}[4.00, 9.15] & 76 & 65\\
Leave one country out: center and range & 5.54 & {}[5.51, 5.58] & 76 & 3009\\
\bottomrule
\end{tabular}
\endgroup{}

}

\end{table}%

\begin{figure}[H]

\centering{

\includegraphics[width=0.95\linewidth,height=\textheight,keepaspectratio]{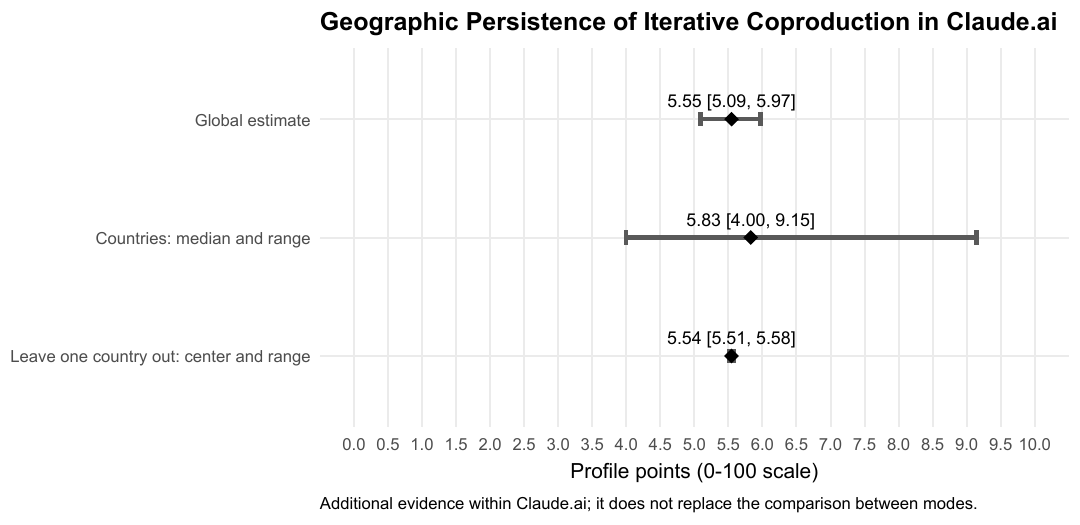}

}

\caption{\label{fig-s6-geo}Geographic persistence within Claude.ai.}

\end{figure}%

\break

\section*{Appendix S7. Observational Limits and
Reproducibility}\label{appendix-s7}
\addcontentsline{toc}{section}{Appendix S7. Observational Limits and
Reproducibility}

Claude.ai can preserve multi-turn conversations. In the 1P API, each
record analyzed by Anthropic is an input-response pair that may belong
to a longer session, but calls do not include metadata linking them to
earlier exchanges (Appel et al. 2026). An iterative intervention
performed through external orchestration may therefore remain outside
the observed unit. The comparison describes profiles constructed
consistently from different telemetry; it does not equate complete
processes.

The cells are monthly aggregates. They do not identify people,
conversations, organizations, or acceptance decisions. Nor do they
observe understanding, epistemic ownership, or individual effort
reallocation. Measurement identifies consistency between traces and
theoretical configurations. Inference concerns aggregate associations,
not causal effects.

The replication package links tables and figures to their generating
objects and reports the bootstrap seed, replications, and support. The
original data sources remain unchanged by the reproduction pipeline.

\section*{References}\label{references}
\addcontentsline{toc}{section}{References}

\phantomsection\label{refs}

\end{document}